# Systematic identification of abundant A-to-I editing sites in the human transcriptome

Erez Y. Levanon[1,2,*], Eli Eisenberg[1,*], Rodrigo Yelin[1,*], Sergey Nemzer[1,*], Martina Hallegger[3], Ronen Shemesh[1], Zipora Y. Fligelman[1], Avi Shoshan[1], Sarah R. Pollock[1], Dan Sztybel[1], Moshe Olshansky[1], Gideon Rechavi[2] & Michael F. Jantsch[3].

[1] *Compugen Ltd., 72 Pinchas Rosen St., Tel-Aviv 69512, Israel*

[2] *Department of Pediatric Hematology-Oncology, Chaim Sheba Medical Center and Sackler School of Medicine, Tel Aviv University, Tel Aviv 52621, Israel*

[3] *Max F. Perutz Laboratories, Dept. of Cell Biology and Genetics, University of Vienna, Rennweg 14, A-1030 Vienna, Austria*

[*] These authors contributed equally to this work.

Correspondence should be addressed to E.Y.L. (e-mail: erez@compugen.co.il).

**Abstract:**

**RNA editing by members of the double-stranded RNA-specific ADAR family leads to site-specific conversion of adenosine to inosine (A-to-I) in precursor messenger RNAs. Editing by ADARs is believed to occur in all metazoa, and is essential for mammalian development. Currently, only a limited number of human ADAR substrates are known, while indirect evidence suggests a substantial fraction of all pre-mRNAs being affected. Here we describe a computational search for ADAR editing sites in the human transcriptome, using millions of available expressed sequences. 12,723 A-to-I editing sites were mapped in 1,637 different genes, with an estimated accuracy of 95%, raising the number of known editing sites by two orders of magnitude. We experimentally validated our method by verifying the occurrence of editing in 26 novel substrates. A-to-I editing in humans primarily occurs in non-coding regions of the RNA, typically in Alu repeats. Analysis of the large set of editing sites indicates the role of editing in controlling dsRNA stability.**

RNA editing by members of the double-stranded RNA-specific ADAR family leads to site-specific conversion of adenosine to inosine (A-to-I) in precursor messenger RNAs[1]. ADAR-mediated RNA editing is essential for normal life and development in both invertebrates and vertebrates[2-5]. ADAR-deficient inverterbrates show only behavioural defects[2, 3], while ADAR1 knock-out mice die embryonically and ADAR2 null mice live to term but die prematurely[4, 5]. High editing levels were found in inflamed tissues[6], in agreement with a proposed antiviral function of ADARs and their transcriptional regulation by interferon[7]. Altered editing patterns were found in epileptic mice[8], suicide victims suffering chronic depression[9], amyotrophic lateral sclerosis[10] and in malignant gliomas[11]. Until recently only a handful of edited human genes were documented, most of which were discovered serendipitously[12]. A systematic experimental search for inosine-containing RNAs has yielded 19 additional cases[13], and one further example was found using a cross-genome comparison approach[14]. However, quantitation of inosine in total RNA suggests that editing affects a much larger fraction of the mammalian transcriptome[15]. In addition, tantalizing hints for abundant editing were observed in high-throughput cDNA sequencing data[16].

Large-scale identification of editing substrates by bioinformatics tools was previously considered practically impossible[17]. In principle, editing may be detected using the large-scale database of ESTs[18] (expressed sequence tags) and RNAs, which currently holds over 5 million human records. Editing sites show up when a sequence is aligned with the genome: while the DNA reads A, sequencing identifies the inosine in the edited site as guanosine (G). However, the poor sequencing quality of the sequence database (up to 3% sequencing errors[19]) precludes a straightforward application of this approach. Moreover, millions of single nucleotide polymorphisms (SNPs) and mutations are erroneously identified as editing events by this method.

Here we present a computational approach that overcomes these challenges. 12,723 A-to-I editing sites were mapped in 1,637 different genes, with an estimated accuracy of 95%. We thus raise the number of known editing sites by two orders of magnitude. Editing was experimentally validated in 26 of these 1,637 genes. The editing sites found are typically located with Alu elements residing in non-coding regions of the RNA. The effect of editing on dsRNA stability is analyzed.

# Results

## Computational identification of A-to-I editing

ADAR substrates are usually imperfect dsRNA stems formed by base pairing of an exon containing the adenosine to be edited with a complementary portion of the pre-mRNA (up to several thousand nucleotides apart). We therefore restricted the search for mismatches to potential double-stranded regions, in order to remove most of the noise and facilitate the identification of true editing sites. For this purpose, human ESTs and cDNAs were aligned to the genome and assembled into clusters representing genes or partial genes. Details of this procedure are given in Sorek et al[20]. Then, our algorithm aligned the expressed part of the gene with the corresponding genomic region, looking for reverse complement alignments longer than 32 bp with identity levels higher than 85% (**Fig. 1**). About 429,000 candidate dsRNAs were found in 14,512 different genes, mostly resulting from alignment of an exon (including the 3` and 5` UTRs) to an intron. In order to further decrease the number of random mismatches, SNPs and mutations, the algorithm then cleaned the sequences supporting the stem region. Since sequencing errors tend to cluster in certain regions, especially in low complexity areas and towards sequences ends, we discarded all single-letter repeats longer than 4 bp, as well as 150 bp at both ends of each sequence. In addition, all 50 nucleotides-wide windows in which the total number of mismatches is 5 or more were considered as having low sequencing quality and were discarded. However, 4 or more identical sequential mismatches were masked in the count for mismatches in a given window. This exception is intended to retain sequences with many sequential editing sites, which were found to occur in previously documented examples[21]. Mismatches supported by less than 5% of available sequences were also discarded, and, finally, known SNPs of genomic origin were removed. Employing these criteria one finds that the putative editing sites tend to group together[16], a fact which is also supported by the few available known cases[13]. Thus, all mismatches that occur less than three times in an exon were ignored.

This above cleaning procedure resulted almost exclusively in A-to-G mismatches (**Fig. 2a**). Employing this procedure we identified 12,723 putative editing sites, belonging to 1,637 different genes. The same approach applied to G-to-A mismatches yielded only 242 sites. Sequencing errors, SNPs and mutations, the three main sources of noise in our analysis, are expected to produce at least as many G-to-As as A-to-Gs (**Fig. 2a**,**b**,**c**). This signal-to-noise ratio (242/12723) suggests that our false positive rate is very low.

## Experimental Validation

To experimentally validate the predicted editing sites, we chose 30 genes and sequenced matching DNA and RNA samples retrieved from the same specimen, for up to five tissues. We have positively verified editing events in 26 previously unknown editing substrates. PCR products were either cloned followed by sequencing of individual clones, or sequenced as a population, without cloning. When the PCR products were cloned, editing occurrence was detected by comparing the sequences of several clones with the genomic sequence (**Fig. 3** and **Supplementary Fig. 1**). When PCR products were directly sequenced, the occurrence of editing was determined by the presence of an unambiguous trace of guanosine in positions for which the genomic DNA clearly indicated the presence of an adenosine (**Fig. 4**). The full sequencing data are given as **Supplementary Figure 2**. We show here, for the first time, direct evidence for editing in liver, lung, kidney, prostate, colon, and uterus. For most genes, editing was found in all tissues, with varying relative abundance, but generally the unedited signal dominated the edited signal. Two genes were validated using cell-lines known to have varying levels of ADAR activity (**Fig. 3**). Interestingly, the observed levels of A-to-I conversions correlated well with the reported ADAR activities in these cell-lines[22]. Typically, additional editing sites, not present in our list, were found in the same region. The validation set was composed of two subsets: (i) 20 genes for which the EST data suggested many putative editing events, 18 of these genes were confirmed to be edited. (ii) 17 genes were chosen randomly from the list of 1,637 predicted genes. Four of these genes were discarded, as they did not allow for designing high-quality amplification primers outside the Alu sequence (see methods). 9 of the remaining 13 genes were successfully amplified and sequenced, 8 of which exhibited editing. Note that the success rate in our random subset (89%) is a lower bound to the true accuracy of the list, as either low editing efficiency at a given site or limited variety of tissues in our validation experiment could prevent the detection of editing events in the experimental sample.

## Characterization of the editing sites

Interestingly, 92% of sites occur within an Alu repeat, and additional 1.3% lie within the primate-L1 repeat, in accordance with previous reports[13, 16]. This is explicable by the fact that only long paired RNA molecules were scanned for editing, a structure more likely formed between repetitive elements. The distribution of editing sites within the Alu sequence exhibits a number of preferred edited adenosines, as well as adenosines unlikely to be edited. In particular, two specific A sites within the Alu repeat, in positions 27 and 28 of Alu, account for ~12% of all editing events (see **Supplementary Note**).

We have also found that G is underrepresented in the nucleotide upstream to the edited A, and overrepresented in the nucleotide following the editing site (see **Supplementary Note**), in accordance with previous reports[23, 24]. However, the fact that most of the sites occur within Alu repeats strongly biases the identification of additional significant patterns characterizing the editing site.

Typically, editing is seen in only a fraction of the supporting expressed sequences (ESTs or cDNAs). In fact, for 83% of the sites only one sequence exhibits editing (after applying our cleaning filters). This suggests that editing does not occur with equal frequency in all tissues and conditions, and is of probabilistic nature. Our experimental data also support this finding.

No specific expression pattern or Gene Ontology (GO) classification for the edited genes was found. However, we analysed the EST libraries searching for specific libraries showing an altered editing pattern. The libraries with the most significant over-editing pattern came from thymus, brain, pancreas, spleen and prostate (see **Supplementary Note**). Some of these observations support previous reports[15, 25].

Editing can extend the proteomic diversity by changing the identity of a particular codon[26], as the ribosome reads inosine as guanosine. Two novel examples of such editing are presented in the **Supplementary Note**. However, Morse et al. have predicted that most pre-mRNA editing in the brain is located in non-coding regions[21]. In agreement with this, virtually all of the editing sites identified by us are located in non-coding regions: of the sites that can be aligned with RefSeq sequences, 12% were located in the 5' UTR, 54% in the 3' UTR and 33% are in RefSeq introns. Some of the sites annotated as introns might actually be within an alternative exon not covered by the RefSeq sequence. Note that our strict cleaning procedure definitely misses many true editing sites. In particular, the known examples of A-to-I editing in the Glutamate receptor and Serotonin receptor were not picked up by our algorithm, as the expressed part of the dsRNA supporting them were not long enough. Thus, it is likely that there are more editing sites within the coding region not detected in this work.

It was suggested that one of the functions of RNA editing is the destabilization of dsRNAs. Our large database of editing sites enables us to test this prediction. ADAR-mediated editing of an A in an A-U base pair produces the less stable I-U pair, while A-C mismatches can be edited into the more stable I-C pairs. Looking at the best complementary alignment of the editing regions, we find that in 78% of the editing cases an A-U pair is destabilized, while in 19% an A-C pair is stabilized. Editing of either A-A or A-G pairs occurs in only 3% of the cases. This suggests that editing is aimed at stabilization and destabilization only, and does not occur in situations where it has no major effect on dsRNA stability. Furthermore, the editing mechanism seems to prefer stabilization: 22% of the editing events target a mismatched base-pair, while the average frequency of such mismatched base-pairs in the sites adjacent to the editing sites is only 10%, since these sites are all located in double-stranded regions. Thus, while most editing events result in destabilization of the dsRNA, we find many more stabilization events (i.e., editing of A-C to I-C) than what would be expected based on a random choice of the editing sites along the dsRNA. The preference towards stabilization editing is in agreement with previous reports[27].

## Discussion

This work increases the number of editing substrates by two orders of magnitude, in accordance with prior estimates[15]. This allows a large-scale analysis of the editing phenomenon. The widespread occurrence of editing makes it a significant contributor to the diversity of the transcriptome, producing presumably more different transcripts than produced by alternative splicing, while affecting only a small number of nucleotides. Interestingly, the large-scale editing in human is found to be strongly associated with Alu repeats, which are unique to primates. Thus, one does not expect the corresponding sites to be found in non-primate mammals. However, other repeats present in these organisms may be associated with the same phenomenon. The pronounced concentration of editing sites in Alu repeats raises that question whether A-to-I editing acts as an anti-transposition mechanism by inhibiting the integration of transcribed Alu back into the genome. Such a scenario is in agreement with an anti-viral mechanism of editing[28], as retrotransposition of many repetitive elements is very similar to some stages of the retroviral infection.

Alternatively, Tonkin et al. suggest[29] that editing regulates RNAi by protecting the dsRNA from degradation. Our results indicate that these possible mechanisms may be of wide applicability. Finally, we note that there are probably many more sites than those listed in this work, since: (i) Editing happens in only a fraction of the sequences. Since the expressed sequence coverage of many genes is scarce, many editing sites might be absent from GenBank sequences. (ii) Our filtering parameters were chosen to minimize the noise, but inevitably miss many true sites such as the known sites in the glutamate receptor and serotonin 2c receptor pre-mRNAs[9, 26]. (iii) The experimental evidence show that a typical editing substrate contains more editing sites than the number predicted by us. Thus, the 12,723 sites we listed may still represent only a portion of the actual editing repertoire. The large-scale mapping of editing sites enables the identification of new properties of non-coding regions, and may facilitate the association of mutations in these regions with known pathologies.

# Methods

## Alignment of expressed sequences to the Genome

Human ESTs and cDNAs were obtained from NCBI GenBank version 136 (June 2003; www.ncbi.nlm.nih.gov/dbEST). The genomic sequences were taken from the human genome build 33 (June 2003; www.ncbi.nlm.nih.gov/genome/guide/human).

Briefly, sequences were aligned as follows: sequences were cleaned from terminal vector sequences, and low-complexity stretches and repeats (including Alu repeats) in the expressed sequences were masked. Then, expressed sequences were compared with the genome to find likely high-quality hits. They were then aligned to the genome by use of a spliced alignment model that allows long gaps. Only sequences having > 94% identity to a stretch in the genome were used in further stages. Further details can be found in Sorek at al[20] .

## Experimental Methods

Total RNA and genomic DNA (gDNA) isolated simultaneously from the same tissue sample were purchased from Biochain Institute. In this work we used samples of liver, prostate, uterus, kidney, colon, lung normal and tumor, brain tumor (glioma), cerebellum and frontal lobe.

The total RNA underwent oligo-dT primed reverse transcription using Superscript II (Invitrogen, Carlsbad, CA) according to manufacturer instructions. The cDNA and gDNA (at 0.1 μg/μl) were used as templates for PCR reactions. We aimed at high sequencing quality and thus amplified rather short genomic sequences (roughly 200 bp). The amplified regions chosen for validation were selected only if the fragment to be amplified maps to the genome at a single site. PCR reactions were done using TaKaRa Ex Taq™ Hot Start (Takara Bio) using the primers and annealing conditions as detailed in the **Supplementary Methods**. The PCR products were run on 2% agarose gels and only if a single clear band of the correct approximate size was obtained, it was excised and sent to Hy-labs laboratories for purification and direct sequencing without cloning.

Poly-A RNA from tissue culture cells was isolated using Trifast (PeqLab) and poly-A selected using magnetic oligo dT beads (Dynal). 1µg of poly A RNA was reverse transcribed using random hexamers as primers and RNAseH deficient M-MLV reverse transcriptase (Promega). Genomic DNA from tissue culture cells was isolated according to Ausubel et al[30].

First strand cDNAs or corresponding genomic regions were amplified with suitable primers using Pfu polymerase, to minimize mutation rates during amplification. Amplified fragments were A-tailed using Taq polymerase, gel purified and cloned into pGem-T easy (Promega). After transformation in E. coli individual plasmids were sequenced and aligned using ClustalW.

We used Contig Express software from Vector NTI 6.0 Suite (Informax, Inc.) for multiple-alignment of the electropherograms (see **Supplementary Fig. 2**). Typically, the extent of A-I editing is variable, e.g. the levels of the guanosine trace sometimes is only a fraction of the adenine trace, while in some occasions the conversion from A to I is almost complete. For each gene tested, we sequenced the three tissues in which the expression was the highest. The RT-PCR and gDNA-PCR of one of these tissues were sequenced from both ends to ensure the consistency of the resulting electropherograms.

Information concerning the editing sites is available in:

http://cgen.com/research/Publications/AtoIEditing

We thank A. Diber, E. Shuster and S. Zevin for technical help. P. Akiva, A. Amit and R. Sorek are acknowledged for critical reading of the manuscript. The work of E.Y.L. was performed in partial fulfilment of the requirements for a Ph.D. degree from the Sackler Faculty of Medicine, Tel Aviv University, Israel. Part of this work was supported by the Austrian Science Foundation grant SFB1706 to MJ.

Figure 1:ADAR-mediated editing: a. pre-mRNA as transcribed from DNA. The gene contains two Alu repeats with opposite orientations, one of which overlaps with an exon. b The two oppositely oriented Alu sequences form a dsRNA structure. c An enzyme of the ADAR family edits some of the adenosines in the dsRNA structure into inosines.

Figure 2: Distribution of mismatches between the DNA and the expressed RNA sequences that pass the cleaning algorithm. a: results of algorithm application to dsRNAs only. A-to-G mismatches clearly dominate the distribution. Notably, T-to-C mismatches are also overrepresented, likely being A-to-I editing events that were aligned to the opposite strand. Inset b shows the distribution of mismatches resulting from applying the algorithm to random expressed sequences covering about 20% of the transcriptome. Insets c and d show the distributions for known SNPs[31] and mutations[32], respectively. A-to-G mismatches do not stand out in the distributions b-d.

Figure 3: Editing in the CFLAR transcript. A region corresponding to the 3' UTR of CFLAR was amplified from cDNAs and gDNA of neuroblastoma, HeLa and HeK293 cells. (A) schematic organization of CFLAR with predicted editing in the 3'UTR (brown shading). There are dozens of Alu elements within the genomic region of the CFLAR gene, and we can not tell for sure which one pairs with the above 3` UTR region (marked with a red arrow) to form the dsRNA required for editing. The closest Alu element is located on the 3` UTR as well, 1450 bp downstream (marked with a blue arrow). For this dsRNA,

virtually all editing events recorded in this figure result in destabilization of the dsRNA. (B) Sequences of individually cloned fragments were aligned to the published human genomic sequence. No A-to-I (reads as G in the sequence) conversion is found in HeK293 cells, while abundant and moderate editing is seen in neuroblastomas and HeLas, respectively. Editing events are highlighted in light brown shading. Nucleotides are numbered according to their position on chromosome 2. An additional example is provided in **Supplementary Figure 1**.

Figure 4: Editing in the F11 receptor (JAM1) gene. Top: some of the publicly available expressed sequences covering this gene, together with the corresponding genomic sequence. The evidence for editing is highlighted. Bottom: Results of sequencing experiments. Matching DNA and cDNA RNA sequences for a number of tissues. Editing is characterized by a trace of guanosine in the cDNA RNA sequence, where the DNA sequence exhibits only adenosine signals (highlighted). 23 additional examples are provided in **Supplementary Figure 2**. Note the variety of tissues showing editing, and the variance in the relative intensity of the edited guanosine signal.

## Reference list

1. Polson, A.G., Crain, P.F., Pomerantz, S.C., McCloskey, J.A. & Bass, B.L. The mechanism of adenosine to inosine conversion by the double-stranded RNA unwinding/modifying activity: a high-performance liquid chromatography-mass spectrometry analysis. *Biochemistry* **30**, 11507–11514 (1991).
2. Tonkin, L.A. et al. RNA editing by ADARs is important for normal behavior in Caenorhabditis elegans. *Embo. J.* **21**, 6025–6035 (2002).
3. Palladino, M.J., Keegan, L.P., O'Connell, M.A. & Reenan, R.A. A-to-I pre-mRNA editing in Drosophila is primarily involved in adult nervous system function and integrity. *Cell* **102**, 437–449 (2000).
4. Wang, Q., Khillan, J., Gadue, P. & Nishikura, K. Requirement of the RNA editing deaminase ADAR1 gene for embryonic erythropoiesis. *Science* **290**, 1765–1768 (2000).
5. Higuchi, M. et al. Point mutation in an AMPA receptor gene rescues lethality in mice deficient in the RNA-editing enzyme ADAR2. *Nature* **406**, 78–81 (2000).
6. Yang, J.H. et al. Widespread inosine-containing mRNA in lymphocytes regulated by ADAR1 in response to inflammation. *Immunology* **109**, 15–23 (2003).
7. Patterson, J.B. & Samuel, C.E. Expression and regulation by interferon of a double-stranded-RNA-specific adenosine deaminase from human cells: evidence for two forms of the deaminase. *Mol. Cell. Biol.* **15**, 5376–5388 (1995).
8. Brusa, R. et al. Early-onset epilepsy and postnatal lethality associated with an editing-deficient GluR-B allele in mice. *Science* **270**, 1677–1680 (1995).
9. Gurevich, I. et al. Altered editing of serotonin 2C receptor pre-mRNA in the prefrontal cortex of depressed suicide victims. *Neuron* **34**, 349–356 (2002).
10. Kawahara, Y. et al. Glutamate receptors: RNA editing and death of motor neurons. *Nature* **427**, 801 (2004).
11. Maas, S., Patt, S., Schrey, M. & Rich, A. Underediting of glutamate receptor GluR-B mRNA in malignant gliomas. *Proc. Natl. Acad. Sci. USA* **98**, 14687–14692 (2001).
12. Bass, B.L. RNA editing by adenosine deaminases that act on RNA. *Annu. Rev. Biochem.* **71**, 817–846 (2002).
13. Morse, D.P. & Bass, B.L. Long RNA hairpins that contain inosine are present in Caenorhabditis elegans poly(A)+ RNA. *Proc. Natl. Acad. Sci. USA* **96**, 6048–6053 (1999).
14. Hoopengardner, B., Bhalla, T., Staber, C. & Reenan, R. Nervous system targets of RNA editing identified by comparative genomics. *Science* **301**, 832–836 (2003).
15. Paul, M.S. & Bass, B.L. Inosine exists in mRNA at tissue-specific levels and is most abundant in brain mRNA. *Embo. J.* **17**, 1120–1127 (1998).
16. Kikuno, R., Nagase, T., Waki, M. & Ohara, O. HUGE: a database for human large proteins identified in the Kazusa cDNA sequencing project. *Nucleic. Acids. Res.* **30**, 166–168 (2002).

17. Seeburg, P.H. A-to-I editing: new and old sites, functions and speculations. *Neuron* **35**, 17–20 (2002).
18. Boguski, M.S., Lowe, T.M. & Tolstoshev, C.M. dbEST--database for "expressed sequence tags". *Nat. Genet.* **4**, 332–333 (1993).
19. Hillier, L.D. et al. Generation and analysis of 280,000 human expressed sequence tags. *Genome Res.* **6**, 807–828 (1996).
20. Sorek, R., Ast, G. & Graur, D. Alu-containing exons are alternatively spliced. *Genome Res.* **12**, 1060–1067 (2002).
21. Morse, D.P., Aruscavage, P.J. & Bass, B.L. RNA hairpins in noncoding regions of human brain and Caenorhabditis elegans mRNA are edited by adenosine deaminases that act on RNA. *Proc. Natl. Acad. Sci. USA* **99**, 7906–7911 (2002).
22. Maas, S. et al. Structural requirements for RNA editing in glutamate receptor pre-mRNAs by recombinant double-stranded RNA adenosine deaminase. *J. Biol. Chem.* **271**, 12221–12226 (1996).
23. Polson, A.G. & Bass, B.L. Preferential selection of adenosines for modification by double-stranded RNA adenosine deaminase. *Embo. J.* **13**, 5701–5711 (1994).
24. Lehmann, K.A. & Bass, B.L. Double-stranded RNA adenosine deaminases ADAR1 and ADAR2 have overlapping specificities. *Biochemistry* **39**, 12875–12884 (2000).
25. Kim, U., Wang, Y., Sanford, T., Zeng, Y. & Nishikura, K. Molecular cloning of cDNA for double-stranded RNA adenosine deaminase, a candidate enzyme for nuclear RNA editing. *Proc. Natl. Acad. Sci. USA* **91**, 11457–11461 (1994).
26. Higuchi, M. et al. RNA editing of AMPA receptor subunit GluR-B: a base-paired intron-exon structure determines position and efficiency. *Cell* **75**, 1361–1370 (1993).
27. Wong, S.K., Sato, S. & Lazinski, D.W. Substrate recognition by ADAR1 and ADAR2. *Rna* **7**, 846–858 (2001).
28. Lei, M., Liu, Y. & Samuel, C.E. Adenovirus VAI RNA antagonizes the RNA-editing activity of the ADAR adenosine deaminase. *Virology* **245**, 188–196 (1998).
29. Tonkin, L.A. & Bass, B.L. Mutations in RNAi rescue aberrant chemotaxis of ADAR mutants. *Science* **302**, 1725 (2003).
30. Ausubel, F.M. et al. Current protocols in molecular biology. (John Wiley & Sons, Inc, New York; 1987).
31. Jiang, R. et al. Genome-wide evaluation of the public SNP databases. *Pharmacogenomics* **4**, 779–789 (2003).
32. Antonarakis, S.E., Krawczak, M. & Cooper, D.C. in The Genetic Basis of Human Cancer, Edn. 2. (eds. B. Vogelstein & K.W. Kinzler) 7–41 (McGraw-Hill, New-York; 2002).

## Supplementary information

### 1. EST libraries

We list here some EST libraries in which the fraction of ESTs showing RNA editing is significantly higher than the average. First, we count all ESTs that are edited at one or more sites out of the 12,723 sites in our database, and compare this number to the total number of ESTs covering these sites that do not exhibit editing (after the cleaning procedure is applied). We find that 6690 ESTs are edited and 4657 are not, giving an average editing to non-editing ratio of 6690:4657 or about 3:2. For each library we then calculate this ratio separately. We list here the libraries most significantly deviating from the 3:2 ratio (p-value calculated by the Fisher's Exact Test).

| Library name | P-value | Number of edited ESTs | Number of ESTs not exhibiting editing | tissue |
|---|---|---|---|---|
| pBluescriptII SK plus | 1.4e-9 | 60 | 5 | Brain |
| NIH_MGC_95 | 1.0$^{e}$-9 | 74 | 9 | Hyppocampus |
| SPLEN2 | 5.6e-7 | 32 | 1 | Spleen |
| THYMU2 | 3.7e-7 | 37 | 2 | Thymus |
| NIH_MGC_110 | 2.2e-7 | 58 | 8 | Pancreas |
| BRACE2 | 1.3e-7 | 43 | 3 | Cerebellum |
| NIH_MGC_83 | 5.5e-5 | 44 | 8 | Prostate |
| CTONG2 | 5.0e-5 | 23 | 1 | Tongue |
| TRACH2 | 4.0e-5 | 19 | 0 | Trachea |
| Stratagene NT2 neuronal precursor 937230 | 4.0$^{e}$-5 | 19 | 0 | Brain |
| BRAMY2 | 1.7$^{e}$-5 | 29 | 2 | Amygdala |
| NIH_MGC_41 | 1.4$^{e}$-5 | 21 | 0 | Cancer, skin |

## 2. Nucleotides distribution

In the following we look at the effect of RNA editing on the stability of its dsRNA substrates. For each predicted site, we search for its best opposite-strand alignment within the genomic region covered by the same gene cluster, and look at the effect of the editing on this alignment. First, we calculate the fraction of editing sites which are (before editing) matching to their opposite strand sequence. We find that 78.3% of the nucleotides in the editing sites match the opposite strand, and 21.7% are mismatched. This frequency of mismatches is actually much higher than could be expected by chance, given that the editing region as a whole is matched with average identity level of about 90%. Indeed, the same analysis for the neighboring sites yields only 10.9% mismatches for the site upstream to the editing site, and 8.3% mismatches for the site downstream to the editing site. Thus the number of matching editing sites is actually lower than expected assuming a uniform distribution of the editing sites on the double-stranded regions.

Next, we look at the distribution of nucleotides in the sites neighboring the editing sites, as well as the site located at the editing sites on the other strand. The distributions are presented in the following table:

| | A | C | G | U |
|---|---|---|---|---|
| Upstream site | 29% | 36% | 6% | 29% |
| Downstream site | 21% | 22% | 43% | 14% |
| Same site opposite strand | 1.5% | 18.9% | 1.3% | 78.3% |

G is strongly underrepresented in the upstream preceding site, and overrepresented in the site following the editing site. However, one should be cautious in analyzing these patterns, as almost all sites are located within highly similar ALU repeats. The site opposed to the editing site is in most cases U, where editing changes the stable A-U pair into the less stable I-U pair. Among the cases in which the edited site is mismatched, the vast majority are C sites, where editing changes the less stable A-C pair into the more stable I-C pair. Changes that do not have a significant effect on the dsRNA stability, i.e., change of A-A pairs into I-A pairs or change of A-G pairs into I-G pairs are rare. This suggests editing is directed at regulating the dsRNA stability. Moreover, the strong bias towards mismatches in the editing sites suggests editing is preferred where it stabilizes the dsRNA.

### 3. Editing sites and the ALU sequence

ALU is a complex and diverse family of genomic repeats that are unique to the primates. Due to their ubiquity, it is probable that two oppositely oriented ALUs will be present in the same gene, and thus they are likely to form dsRNAs and putative editing sites. We thus compared our editing sites with the ALU repeat. In order to simplify the following analysis we concentrated on a "generic" ALU consensus sequence. We used the consensus of the Alu-J subfamily. The exact sequence that was used is gnl|alu|HSU14567.

All 12,723 predicted editing sites were compared to the ALU sequence using the BLASTN program. The best same-strand hits to ALU were used. More than 93% of the sequences in the database had a significant (E-score <1e-10) match to the ALU consensus sequence. We retained only hits with at least 80% identity, which contain the predicted edited site in the alignment. We found 10,928 such hits, each assigning one editing site to a specific position on the ALU sequence.

The ALU consensus sequence is 290 nt in length, and contains 67 A's (23.1% of sequence). Of the 10,928 counts of predicted editing positions with alignments to ALU, 6,861 (63%) are in A positions. The remaining sites are almost exclusively located in G positions (i.e., a site which corresponds to a G in the ALU but actually shows A in the DNA, is edited to be G). This and more information is summarized in the following table:

| Nucleotide | Absolute frequency in ALU sequence | Percentage in ALU sequence | Number of predicted editing sites aligned to this nucleotide | Percentage of predicted editing sites aligned to this nucleotide |
|---|---|---|---|---|
| A | 67 | 23.1% | 6,861 | 63% |
| C | 82 | 28.3% | 157 | 1.4% |
| G | 97 | 33.5% | 3588 | 33% |
| T | 44 | 15.2% | 293 | 2.7% |

In the following table we present the distribution of the 3,802 edited sites located in A positions and aligned to the Alu-J sequence, excluding reverse-complement alignments. It is shown that there are preferred positions for editing events in the alignment to ALU (p-value calculated using the Z-test). Note that positions 27,28 and 162 account for 19% of the positions aligned to A. This is a large bias suggesting that these 3 positions are in a place very favorable for RNA editing. In contrast, position 44 and 179 (only a few bases apart) has a counts of just 8 and 5, respectively, showing that these positions is unfavorable for predicted editing. Such very close positions with significantly different counts serve as ideal control for each other as there was no prior selection that preferred any of them. It is also noteworthy that all A’s in the last 60bp of the sequence are significantly underrepresented.

| Position of A in ALU | Counts of predicted editing sites aligned to this position | P-value of count being far from expected count (56.73 with std of 7.53) |
|---|---|---|
| 19 | 78 | 0.002 |
| 27 | 221 | < 1e-15 |
| 28 | 249 | < 1e-15 |
| 33 | 71 | 0.03 |
| 36 | 37 | 0.004 |
| 44 | 8 | 5e-11 |
| 50 | 48 | 0.123 |
| 57 | 3 | 5e-13 |
| 60 | 5 | 3e-12 |
| 63 | 92 | 1e-6 |
| 68 | 73 | 0.015 |
| 73 | 135 | < 1e-15 |
| 76 | 19 | 3e-7 |
| 82 | 93 | 7e-7 |
| 84 | 27 | 4e-5 |
| 87 | 72 | 0.021 |
| 96 | 65 | 0.14 |
| 97 | 160 | < 1e-15 |
| 99 | 70 | 0.04 |
| 101 | 85 | 9e-5 |
| 105 | 51 | 0.22 |
| 106 | 68 | 0.067 |
| 107 | 133 | < 1e-15 |
| 118 | 110 | 8e-13 |
| 120 | 11 | 6e-10 |
| 121 | 114 | 1e-14 |
| 122 | 76 | 0.005 |
| 123 | 80 | 0.001 |
| 124 | 69 | 0.052 |
| 125 | 51 | 0.223 |
| 127 | 127 | < 1e-15 |
| 129 | 19 | 3e-7 |
| 130 | 52 | 0.26 |
| 131 | 41 | 0.018 |
| 132 | 78 | 0.002 |
| 133 | 43 | 0.034 |
| 136 | 181 | < 1e-15 |
| 162 | 248 | < 1e-15 |
| 168 | 51 | 0.22 |
| 172 | 105 | 7e-15 |
| 179 | 5 | 3e-12 |
| 185 | 43 | 0.034 |
| 189 | 91 | 3e-06 |
| 192 | 10 | 3e-10 |
| 195 | 49 | 0.15 |
| 203 | 71 | 0.029 |
| 211 | 8 | 5e-11 |
| 217 | 9 | 1e-10 |
| 224 | 48 | 0.123 |
| 228 | 51 | 0.223 |
| 235 | 18 | 1e-7 |
| 243 | 7 | 2e-11 |
| 248 | 17 | 7e-8 |
| 253 | 7 | 2e-11 |
| 263 | 4 | 1e-12 |
| 265 | 14 | 7e-9 |
| 267 | 9 | 1e-10 |
| 271 | 13 | 3e-9 |
| 273 | 5 | 3e-12 |
| 283 | 0 | 2e-14 |
| 284 | 1 | 7e-14 |
| 285 | 1 | 7e-14 |
| 286 | 0 | 2e-14 |
| 287 | 2 | 2e-13 |
| 288 | 0 | 2e-14 |
| 289 | 0 | 2e-14 |
| 290 | 0 | 2e-14 |

## 4. Putative editing sites in CDS

### HSPC274

```
ATGGAGTCTAGTTCTGTCGCCCAGGCTGGAGCCCAGTGGTGTGATCTCAGCTCACTGCAATCTCCACCTCCTGGCTTCAAGCGATTCTTCTGCTTCAGCCTCC genome
ATGGAGTCTAGTTCTGTCGCCCAGGCTGGAGCCCAGTGGTGTG TCTCAGCTCACTGCAATCTCCACCTCCTGGCTTCAAGCGATTCTTCTGCTTCAGCCTCC AK055700
ATGGAGTCTRGTTCTGTCGCCCAGGCTGGAGCCCAGTGGTGTGATCTCAGCTCACTGCAATCTCCACCTCCTGGCTTCAAGCGATTCTTCTGCTTCAGCCTCC AF161392
ATGGAGTCTGGTTCTGTCGCCCAGGCTGGAGCCCAGTGGTGTGATCTCAGCTCACTGCAATCTCCACCTCCTGGCTTCAAGCGATTCTTCTGCTTCAGCCTCC BE866764
ATGGAGTCTAGTTCTGTCGCCCAGGCTGGAGCCCAGTGGTGTGATCTCAGCTCACTGCAATCTCCACCTCCTGGCTTCAAGCGATTCTTCTGCTTCAGCCTCC BG025753
ATGGAGTCTGGTTCTGTCGCCCAGGCTGGAGCCCAGTGGTGTGATCTCAGCTCACTGCAATCTCCGCCTCCTGGCTTCAAGCGATTCTTCTGCTTCAGCCTCC BE392757
ATGGAGTCTAGTTCTGTCGCCCAGGCTGGAGCCCAGTGGTGTGATCTCAGCTCACTGCAATCTCCACCTCCTGGCTTCAAGCGATTCTTCTGCTTCAGCCTCC BI907157
ATGGAGTCTGGTTCTGTCGCCCAGGCTGGAGCCCAGTGGTGTGATCTCGGCTCACTGCAATCTCCGCCTCCTGGCTTCAGGCGATTCTTCTGCTTCAGCCTCC CB999681
ATGGAGTCTGGTTCTGTCGCCCAGGCTGGAGCC AGTGGTGTGATCTCAGCTCACTGCAATCTCCACCTCCTGGCTTCAAGCGATTCTTCTGCTTCAGCCTCC AA158593
ATGGAGTCTGGTTCTGTCGCCCAGGCTGGAGCCCAGTGGTGTGATCTCAGCTCACTGCAATCTCCACCTCCTGGCTTCAAGCGATTCTTCTGCTTCAGCCTCC BU959178
ATGGAGTCTGGTTCTGTCGCCCAGGCTGGAGCCCAGTGGTGTGATCTCAGCTCACTGCAATCTCCACCTCCTGGCTTCAAGCGATTCTTCTGCTTCAGCCTCC BG401333
ATGGAGNNTAGTTCTGTCGCCCAGGCTGGAGCCCAGTGGTGTGATCTCAGCTCACTGCAATCTCCACCTCCTGGCTTNAAGCGATTCTTCTGCTTCAGCCTCC F06429
ATGGAGTCTGGTTCTGTCGCCCAGGCTGGAGCCCAGTGGTGTGATCTCAGCTCACTGCAATCTCCACCTCCTGGCTTCAAGCGATTCTTCTGCTTCAGCCTCC AL551550
ATGGAGTCTGGTTCTGTCGCCCAGGCTGGAGCCCAGTGGTGTGATCTCAGCTCACTGCAATCTCCGCCTCCTGGCTTCAAGCGATTCTTCTGCTTCAGCCTCC BX441040
```

In The HSPC274 (C20orf30) gene, exon 2 is an alternative ALU based exon. The alternative exon contains a few putative editing sites, some of which are silent, but others cause replacement of amino acids: In the strongest site no AA is changed (position 10 CTA>CTG L>L). However, transition of A>G at position 66 replaces His with Arg (CAC>CGC H>R). A Change of an AA is noted also in position 80 (AGC>GGC S>G) (ORF was taken from the AF161392 sequence)

### FLJ25952

```
CTGGCGGATCACCTGAGGTCAGGGGTTCGAGATTAGCCTGGTCAAAATGGCAAAACCCTGTCTCCACTAAAAA     GENOME
```

```
CTGGCGGATCACCTGAGGTCAGGGGTTCGAGATTAGCCTGGTCAAAATGGCAAAACCCTGTCTCCACTAAAAA     AK090979
CTGGCGGATCACCTGAGGTCAGGGGTTCGAGATTAGCCTGGTCAAAATGGCAAAACCCTGTCTCCACTAAAAA     AK098818
CTGGCGGATCACCTGAGGTCAGGGGTTCGAGATTAGCCTGGTCAAAATGGCAAAACCCTGTCTCCACTAAAAA     NM153251
------------------------GTTCGAGATTAGCCTGGTCAAGATGGCAAAACCCTGTCTCCACTAAAAA     W16480
--------TCACCTGAGGTCAGGGGTTCGAGATTAGCCTGGTCAAAATGGCAAAACCCTGTCTCCACTAAAAA     BM556200
CTGGCGGATCACCTGAGGTCAGGGGTTCGAGATTAGCCTGGTCGAAATGGCAAAACCCTGTCTCCACTAAAAA     BF207712
CTGGCGGATCACCTGAGGTCAGGGGTTCGAGATTAGCCTGGTCGGGATGGCAAAACCCTGTCTCCACTAAAAA     CD518207
```

```
TACAAAAAAACCCCAAAACTGTCCAGGCATGGTGGCACACGCCTGTAGTCCCAACTACTCGGGAGGTGGAGGCAGGAG     GENOME
TACAAAAAAACCCCAAAACTGTCCAGGCATGGTGGCACACGCCTGTAGTCCCAGCTACTCGGGAGGTGGAGGCAGGAG     AK090979
TACAAAAAAACCCCAAAACTGTCCAGGCATGGTGGCACACGCCTGTGGTCCCAACTACTCGGGAGGTGGAGGCAGGAG     AK098818
TACAAAAAAACCCCAAAACTGTCCAGGCATGGTGGCACACGCCTGTGGTCCCAACTACTCGGGAGGTGGAGGCAGGAG     NM153251
TACAAAAAAACCCCAAAACTGTCCAGGCATGGTGGCACACGCCTGTGGTCCCAACTACTCGGGAGGTGGAGGCAGGAG     W16480
TACAAAAAAACCCCAAAACTGTCCAGGCATGGTGGCACACGCCTGTAGTCCCAACTACTCGGGAGGTGGAGGCAGGAG     BM556200
TACAAAAAAACCCCAAAACTGTCCAGGCATGGTGGCACACGCCTGTGGTCCCAACTACTCGGGAGGTGGAGGCAGGAG     BF207712
TACAAAAAAACCCCAAAACTGTCCAGGCATGGTGGCACACGCCTGTGGTCCCAGCTGCTCGGGAGGTGGAGGCAGGAG     CD518207
```

The hypothetical protein FLJ25952 contains a few putative editing sites. The sequence in genome positions 44-46 is tcaaaa (SK). Editing changes the sequence in one of a number of ways: tcgaaa (SK), tcaaga(SR), or tcggga(SG).

Other potential editing sites in this exon are: position 120 agt> ggt S>G, position 127 aac > agc N>S, and position 130 tac>tgc Y >C.

**Figure 1**

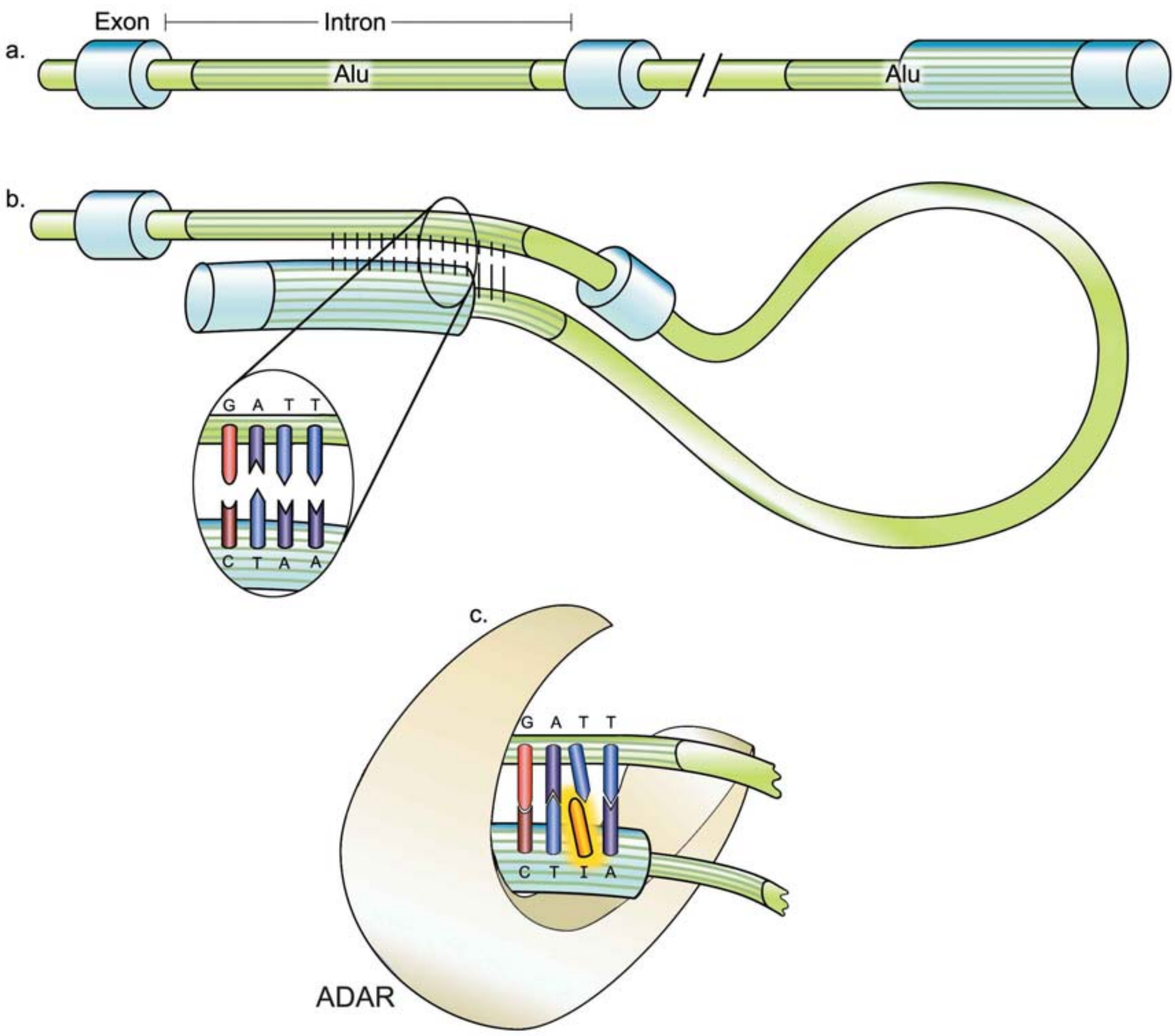

**Figure 2**

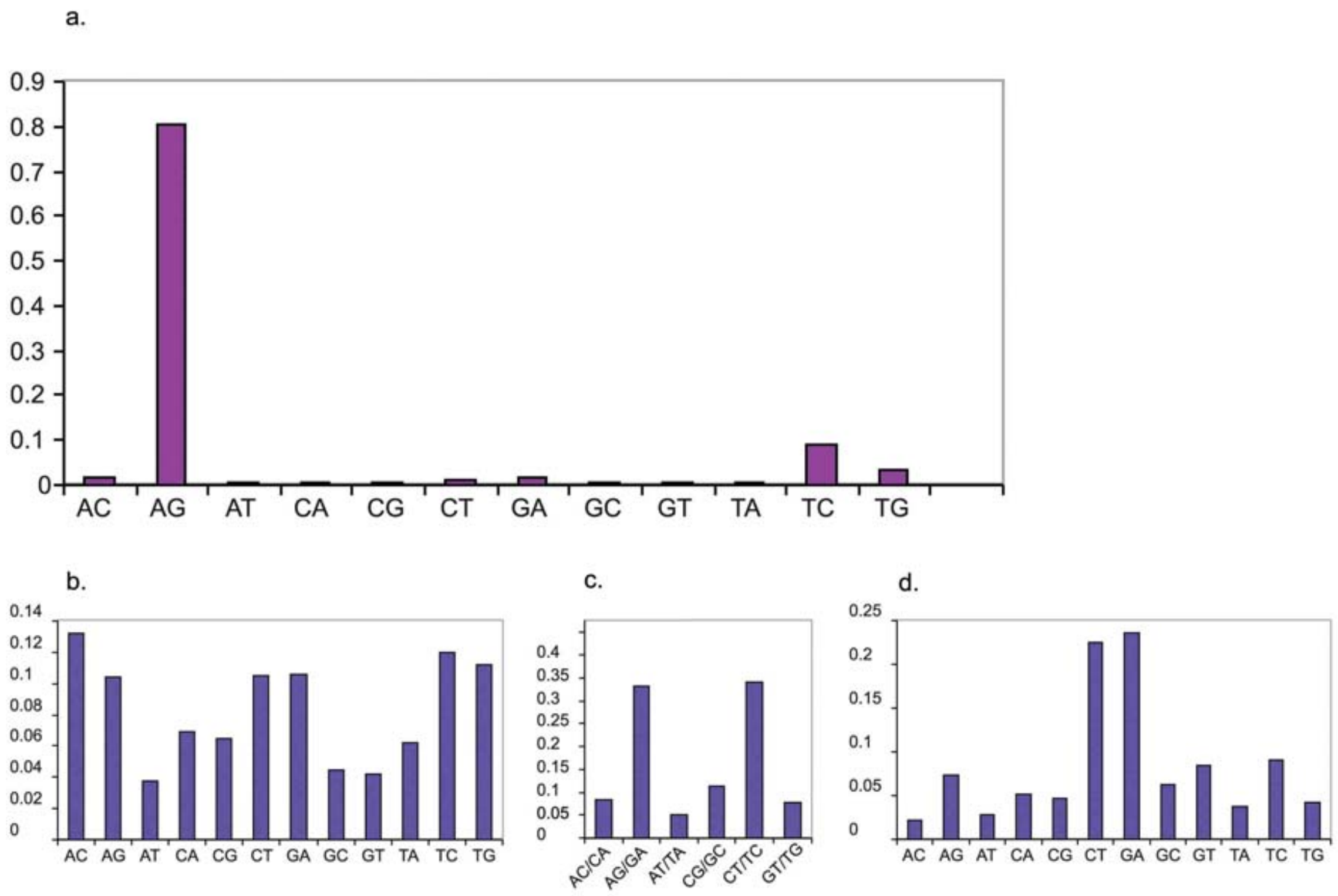
a.
0.9
0.8
0.7
0.6
0.5
0.4
0.3
0.2
0.1
0
AC AG AT CA CG CT GA GC GT TA TC TG
b.
0.14
0.12
0.1
0.08
0.06
0.04
0.02
0
AC AG AT CA CG CT GA GC GT TA TC TG
c.
0.4
0.35
0.3
0.25
0.2
0.15
0.1
0.05
0
AC/CA AG/GA AT/TA CG/GC CT/TC GT/TG
d.
0.25
0.2
0.15
0.1
0.05
0
AC AG AT CA CG CT GA GC GT TA TC TG

Figure 3

**a**

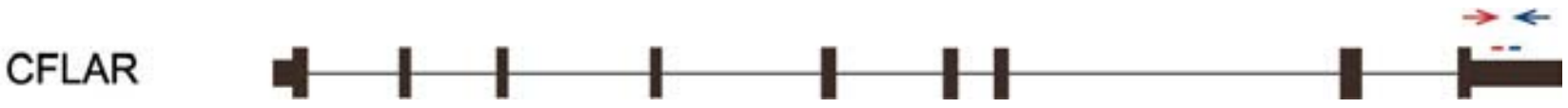

**b**

```
Hek1     GGCTCACACCTGTAATCCCAGCACTTTGGGAGGCCAAGGAGGGCAGATCACTTCAGGTCA
Hek2     GGCTCACACCTGTAATCCCAGCACTTTGGGAGGCCAAGGAGGGCAGATCACTTCAGGTCA
HeLa1    GGCTCACACCTGTAATCCCAGCACTTTGGGAGGCCAAGGAGGGCAGATCACTTCAGGTCA
HeLa2    GGCTCACACCTGTAATCCCAGCACTTTGGGAGGCCAAGGAGGGCAGATCACTTCAGGTCA
HeLa3    GGCTCACACCTGTAATCCCAGCACTTTGGGAGGCCAAGGAGGGCAGATCACTTCAGGTCA
HeLa4    GGCTCACACCTGTAATCCCAGCACTTTGGGAGGCCAAGGAGGGCAGATCACTTCAGGTCA
HeLa5    GGCTCACACCTGTAATCCCAGCACTTTGGGAGGCCAAGGAGGGCAGATCACTTCAGGTCA
HeLa6    GGCTCACACCTGTAATCCCAGCACTTTGGGAGGCCAAGGAGGGCAGATCACTTCAGGTCA
NB1      GGCTCACACCTGTAATCCCAGCACTTTGGGAGGCCAAGGAGGGCAGATCACTTCAGGTCA
NB2      GGCTCACACCTGTAATCCCAGCACTTTGGGAGGCCAAGGAGGGCAGATCACTTCAGGTCA
NB3      GGCTCACACCTGTAATCCCAGCACTTTGGGAGGCCAAGGAGGGCAGATCACTTCAGGTCA
NB4      GGCTCACACCTGTAATCCCAGCACTTTGGGAGGCCAAGGAGGGCAGATCACTTCAGGTCA
NB5      GGCTCACGCCTGTAATCCCAGCACTTTGGGAGGCCGAGGAGGGCAGATCACTTCAGGTCA
NB6      GGCTCACGCCTGTGGTCCCAGCACTTTGGGAGGCCGAGGAGGGCAGATCACTTCAGGTCG
NB7      GGCTCACGCCTGTGGTCCCAGCACTTTGGGAGGCCGAGGAGGGCAGATCACTTCAGGTCG
NB8      GGCTCGCACCTGTGGTCCCAGCACTTTGGGAGGCCGAGGAGGGCAGATCACTTCAGGTCG
NB9      GGCTCGCGCCTGTGGTCCCAGCACTTTGGGAGGCCGAGGGGGGCAGATCACTTCAGGTCG
NB10     GGCTCGCGCCTGTAATCCCAGCACTTTGGGAGGCCAAGGAGGGCAGATCACTTCAGGTCA
NB11     GGCTCGCGCCTGTGGTCCCAGCACTTTGGGAGGCCGAGGAGGGCAGATCACTTCAGGTCA
CHROMO1  GGCTCACACCTGTAATCCCAGCACTTTGGGAGGCCAAGGAGGGCAGGTCACTTCAGGTCA
CHROMO2  GGCTCACACCTGTAATCCCAGCACTTTGGGAGGCCAAGGAGGGCAGATCACTTCAGGTCA
CFLAR    GGCTCACACCTGTAATCCCAGCACTTTGGGAGGCCAAGGAGGGCAGATCACTTCAGGTCA

Hek1     GGAGTTCGAGACCAGCCTGGCCAACATGGTAAACGCTGTCCCTAGTAAAAATACAAAAAT
Hek2     GGAGTTCGAGACCAGCCTGGCCAACATGGTAAACGCTGTCCCTAGTAAAAATACAAAAAT
HeLa1    GGAGCTCGAGACCAGCCTGGCCAACATGGTAAACGCTGTCCCTAGTAAAAATACAAAAAT
HeLa2    GGAGTTCGAGACCAGCCTGGCCAACATGGTAAACGCTGTCCCTAGTAAAAATACAAAAAT
HeLa3    GGAGTTCGAGACCAGCCTGGCCAACATGGTAAACGCTGTCCCTAGTAAAAATGCAGAAAT
HeLa4    GGAGTTCGAGACCAGCCTGGCCAACATGGTAAACGCTGTCCCTAGTAAAAATACAAAAAT
HeLa5    GGAGTTCGAGACCAGCCTGGCCAACATGGTAAACGCTGTCCCTAGTAAAAGTACAAAAAT
HeLa6    GGAGTTCGAGACCAGCCTGGCCAACATGGTAAACGCTGTCCCTAGTAAAAATACAAAAAT
NB1      AGAGTTCGAGACCAGCCTGGCCAACATGGTAAACGCTGTCCCTAGTAAAAATACAAAAAT
NB2      GGAGTTCGAGACCAGCCTGGCCAACATGGTAAACGCTGTCCCTAGTAAAAATACAAAAAT
NB3      GGAGTTCGAGACCAGCCTGGCCAACATGGTAAACGCTGTCCCTAGTAAAAATACAAAAAT
NB4      GGAGTTCGAGACCAGCCTGGCCAACATGGTAAACGCTGTCCCTAGTAAAAATACAAAAAT
NB5      GGAGTTCGAGACCAGCCTGGCCAACATGGTAAACGCTGTCCCTAGTAAAAATGCAAAAAT
NB6      GGGGTTCGAGACCAGCCTGGCCAACATGGTAAACGCTGTCCCTAGTGGAAATGCAGGAAT
NB7      GGGGTTCGAGACCAGCCTGGCCAACATGGTAAACGCTGTCCCTAGTGGAAATGCAGGAAT
NB8      GGAGTTCGAGACCAGCCTGGCCAGCATGGTAGACGCTGTCCCTAGTAGAAATGCAGAGGT
NB9      GGAGTTCGAGACCAGCCTGGCCAACATGGTAGACGCTGTCCCTAGTAGAAGTACAGAAAT
NB10     GGAGTTCGAGACCAGCCTGGCCAACATGGTAAACGCTGTCCCTAGTAAAAATACAAAAAT
NB11     GGGGTTCGAGACCGGCCTGGCCAACATGGTGGACGCTGTCCCTGGTAAAAGTGCAGAAAT
CHROMO1  GGAGTTCGAGACCAGCCTGGCCAACATGGTAAACGCTGTCCCTAGTAAAAATACAAAAAT
CHROMO2  GGAGTTCGAGACCAGCCTGGCCAACATGGTAAACGCTGTCCCTAGTAAAAATACAAAAAT
CFLAR    GGAGTTCGAGACCAGCCTGGCCAACATGGTAAACGCTGTCCCTAGTAAAAATACAAAAAT

Hek1     TAGCTGGGTGTGGGTGTGGGTACCTGTATTCCCAGTTACTTGGGAGGCTGAGGTGGGAGG
Hek2     TAGCTGGGTGTGGGTGTGGGTACCTGTATTCCCAGTTACTTGGGAGGCTGAGGTGGGAGG
HeLa1    TAGCTGGGTGTGGGTGTGGGTACCTGTATTCCCAGTTACTTGGGAGGCTGAGGTGGGAGG
HeLa2    TAGCTGGGTGTGGGTGTGGGTACCTGTATTCCCAGTTACTTGGGAGGCTGAGGTGGGAGG
HeLa3    TAGCTGGGTGTGGGTGTGGGTACCTGTATTCCCAGTTACTTGGGAGGCTGAGGTGGGAGG
HeLa4    TAGCTGGGTGTGGGTGTGGGTACCTGTATTCCCAGTTACTTGGGAGGCTGAGGTGGGAGG
HeLa5    TAGCTGGGTGTGGGTGTGGGTACCTGTATTCCCAGTTACTTGGGAGGCTGAGGTGGGAGG
HeLa6    TAGCTGGGTGTGGGTGTGGGTACCTGTATTCCCAGTTACTTGGGAGGCTGAGGTGGGAGG
NB1      TAGCTGGGTGTGGGTGTGGGTACCTGTATTCCCAGTTACTTGGGAGGCTGAGGTGGGAGG
NB2      TAGCTGGGCGTGGGTGTGGGTACCTGTATTCCCAGTTACTTGGGAGGCTGAGGTGGGAGG
NB3      TAGCTGGGTGTGGGTGTGGGTACCTGTATTCCCAGTTACTTGGGAGGCTGAGGTGGGAGG
NB4      TAGCTGGGTGTGGGTGTGGGTACCTGTATTCCCAGTTACTTGGGAGGCTGAGGTGGGAGG
NB5      TAGCTGGGTGTGGGTGTGGGTACCTGTATTCCCAGTTACTTGGGAGGCTGAGGTGGGAGG
NB6      TGGCTGGGTGTGGGTGTGGGTACCTGTGTTCCCAGTTACTTGGGAGGCTGGGGTGGGGGG
NB7      TGGCTGGGTGTGGGTGTGGGTACCTGTGTTCCCAGTTACTTGGGAGGCTGGGGTGGGGGG
NB8      TAGCTGGGTGTGGGTGTGGGTGCCTGTGTTCCCGGTTGCATGGGAGGCTGAGGTGGGAGG
NB9      TAGCTGGGTGTGGGTGTGGGTACCTGTGTTCCCAGTTACTTGGGAGGCTGAGGTGGGAGG
NB10     TAGCTGGGTGTGGGTGTGGGTACCTGTATTCCCAGTTACTTGGGAGGCTGAGGTGGGAGG
NB11     TGGCTGGGTGTGGGTGTGGGTACCTGTGTTCCCAGTTACTTGGGAGGCTGAGGTGGGAGG
CHROMO1  TAGCTGGGTGTGGGTGTGGGTACCTGTATTCCCAGTTACTTGGGAGGCTGAGGTGGGAGG
CHROMO2  TAGCTGGGTGTGGGTGTGGGTACCTGTATTCCCAGTTACTTGGGAGGCTGAGGTGGGAGG
CFLAR    TAGCTGGGTGTGGGTGTGGGTACCTGTATTCCCAGTTACTTGGGAGGCTGAGGTGGGAGG
```

```
GENOME    CAGGAGTTCAAGATCAGCCTGACCAACATGGAGAAACCCTACTAAAAA
AI093487  CAGGAGTTCGGGATCAGCCTGACCAACATGGAGAAACCCTACTGGGAA
BF771639  CGGGGGTTCGAGATCAGCCTGACCAACATGGAGAAACCCTACTGAAAA
BM681047  CGGGAGTTCGGGATCAGCCTGACCAACATGGAGAAACCCTACTGAAAA
BQ307221  CAGGAGTTCAGGATCAGCCTGACCAACATGGAGAAACCCTACTGNGAA
R01692    CAGGAGTTCAGGATCAGCCTGACCAACATGGAGAAACCCTACTGGAAA
BQ305305  CAGGAGTTCAGGATCAGCCTGACCAACATGGAGAAACCCTACTGGGAA
AA101562  CGGGAGTTCGGGATCAGCCTGACCAACATGGAGAAACCCTACTGGAAA
AW190875  CGGGAGTTCAGGATCAGCCTGACCAACATGGAGAAACCCTACTGGAAA
BE350662  CGGGAGTTCGAGATCAGCCTGACCAACATGGAGAAACCCTGCTGAAAA
AA149993  CGGGAGTTCGGGATCAGCCTGACCAACATGGAGAAACCCTACTGGGAA
AI925871  CAGGAGTTCAAGATCAGCCTGACCAACATGGAGAAACCCTACTAAAAA
AI333843  CGGGAGTTCAGGATCAGCCTGACCAACATGGAGAAACCCTACTGGGAA
AW338261  CAGGAGTTCGGGATCAGCCTGACCAACATGGAGAAACCCTACTGGAAA
```

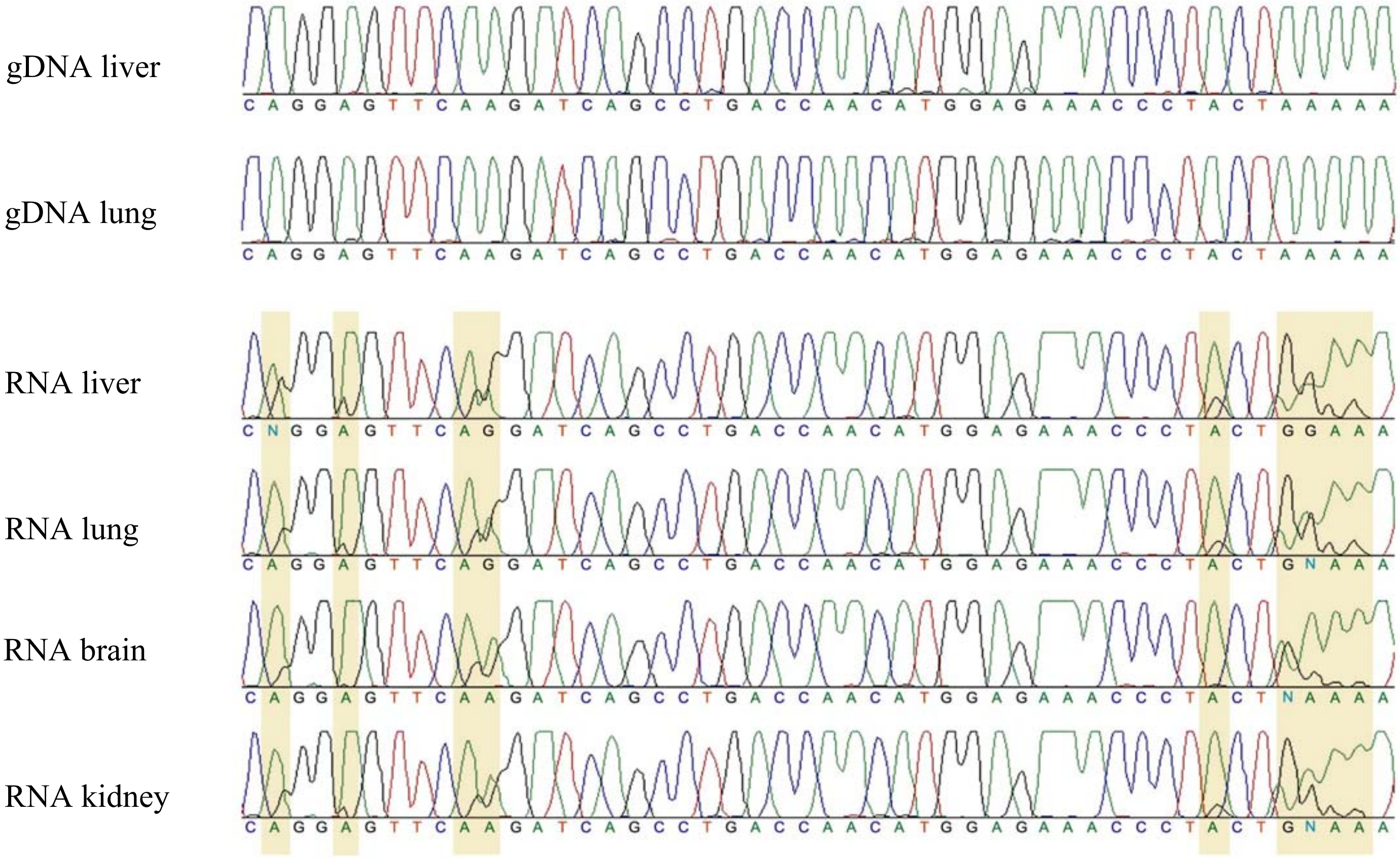